\documentclass[prb,twocolumn,showpacs,amsmath,amssymb,superscriptaddress]{revtex4}
\usepackage{graphicx}
\usepackage{graphics}
\usepackage{dcolumn}
\usepackage{bm}
\begin{document}
\title{Dielectric Function of Diluted Magnetic Semiconductors in the Infrared Regime}
\author{R. Aguado}
\affiliation{Instituto de Ciencia de Materiales de Madrid (CSIC),
Cantoblanco, 28049, Madrid, Spain.}
\author{M.P. L\'opez-Sancho}
\affiliation{Instituto de Ciencia de Materiales de Madrid (CSIC),
Cantoblanco, 28049, Madrid, Spain.}
\author{Jairo Sinova}
\affiliation{Department of Physics, Texas A\&M University, College
Station, TX 77843-4242}
\author{L. Brey}
\affiliation{Instituto de Ciencia de Materiales de Madrid (CSIC),
Cantoblanco, 28049, Madrid, Spain.}
\date{\today}

\begin{abstract}
We present a study of the dielectric function of metallic
(III,Mn)V diluted magnetic semiconductors in the infrared regime.
Our theoretical approach is based on the kinetic exchange model
for carrier induced (III,Mn)V ferromagnetism. The dielectric
function is calculated within the random phase approximation and,
within this metallic regime, we treat disorder effects
perturbatively and thermal effects within the mean field
approximation. We also discuss the implications of this
calculations on carrier concentration measurements from the
optical f-sum rule and the analysis of plasmon-phonon coupled modes
in Raman spectra.
\end{abstract}

\maketitle

\section{Introduction}
Ever since post-growth annealing procedures in (III,Mn)V diluted
magnetic semiconductors (DMS)\cite{Ohno:1996_} demonstrated the
ability to increase the ferromagnetic transition temperature,
$T_c$, by almost two fold, the study of these promising and
interesting materials has been vigourous both experimental and
theoretical. The observed increase in $T_c$ is due to the increase
of the free hole-carrier concentration originally induced by the
Mn substitutional impurities but partially compensated by other
impurities such as  As-antisites and  Mn-interstitials which arise
from the  needed non-equilibrium growth of these materials.
\cite{Ohno:1999_a,Ku:2003_,Yu:2002_a,Edmonds:2004_,Kuryliszyn-Kudelska:2003_cond-mat/0304622}
Because of the sensitivity of their magnetic state to growth
conditions, doping, and external fields, and the strong
valence-band spin-orbit coupling the transport and optical
properties of these heavily-doped semiconductors are richer than
those of conventional itinerant electron ferromagnets.

We present a theory of the infrared dielectric function of
(III,Mn)V ferromagnets based on the kinetic exchange model for
carrier induced (III,Mn)V ferromagnetism which assumes a sallow
acceptor picture. This justifies the treatment of the electronic
structure of the free carriers  by the unperturbed host
semiconductor band structure.  The Mn 3d$^5$ electrons are
strongly localized with a fully polarized $S=5/2$ moment which
interacts with the free carriers via Coulomb and short-range
exchange interactions, \cite{Dietl:2000_,Dietl:2001_b,Konig:2003_}
as has been demonstrated by electron paramagnetic resonance (EPR)
and optical measurements. \cite{Linnarsson:1997_,Ohno:1999_a} As
described in the next section, the dielectric function is
calculated within the random phase approximation and, in this
metallic regime, we treat disorder effects perturbatively and
thermal effects within the mean field approximation.

In particular, we demonstrate that infrared dielectric function
measurements in metallic samples can be used to obtain information
about the carrier concentration, to simulate the plasmon-phonon
overdamped coupled modes, and to explain naturally the observed
optical absorption measurements in the infrared regime arising
from inter-valence-band transitions. Some of our considerations
are based on standard linear-response theory expressions for
weakly disordered metals, in which disorder is included through
finite quasiparticle lifetimes and localization effects (demonstrated
in this model through finite-size exact
diagonalization studies \cite{Yang:2003_}) are taken into account
phenomenologically by weighting differently intra-band and
inter-band quasiparticle lifetimes. We estimate the magnitude of
these lifetimes from extensive theoretical studies in dc-transport
studies which describe successfully the metallic DMS  high-$T_C$
regime.\cite{Jungwirth:2002_c,Lopez-Sancho:2003_}

Although the predictions discussed below are intended to be more
reliable for the most metallic systems\cite{note_dassarma}, they
appear to explain much currently available infrared optical data,
many of which has been obtained in studies of systems with
relatively low dc conductivities.

The paper is organized as follows. In Sec.~\ref{model} we briefly describe
the model Hamiltonian and our theoretical approach and approximations.
In Secs.~\ref{results} we analyze the results of the dielectric function
calculations, sum rules, and plasmon-phonon coupled modes. We then
present a summary in Sec.~\ref{summary}

\section{Model Hamiltonian}
\label{model} Focusing on the infrared regime, in our calculations
we describe the free carriers electronic structure  of the DMS
using the six band Kohn-Luttinger Hamiltonian.  The dielectric
function is calculated in the Random Phase Approximation (RPA)
formalism and finite disorder effects are taken into account
perturbatively by introducing a finite quasiparticle lifetime. In
addition, thermal fluctuations are ignored and finite temperature
effects are only considered at the mean field level. Hence, the
system is described by the following Hamiltonian,
\begin{equation}
H=H_{\rm holes}+H_{\rm Mn^{2+}-holes}+J\sum_{I,i} \mathbf{S} _I \cdot \mathbf{s}_i \, \,
\delta (\mathbf{r}_i -\mathbf{R}_I)  \, \, \, \,  , \label{Htotal}
\end{equation}
Where $H_{holes}$ is the part of the Hamiltonian which describes
the itinerant holes within the ${\bf k}\cdot{\bf p}$ theory,
$H_{\rm Mn^{2+}-holes}$ describes the screened Coulomb
interactions between the holes and localized Mn ions, and the last
term is the antiferromagnetic exchange interaction with strength
$J$ between the spin of the Mn$^{2+}$ ions located at the random
positions $\mathbf{R}_I$ and the spin $\mathbf{s}_i$ of the hole
carriers.

The sources  of disorder  known  to be  relevant in  these
materials include the positional randomness, $\mathbf{R}_I$  of the substitutional Mn ions
with charge $-e$, random placement of interstitial Mn ions, acting
as non-ferromagnetic double donors,\cite{Blinowski:2003_}
and  non-magnetic As anti-sites acting also  as having  charge $+2e$.
These Coulomb interactions are included in $H_{\rm Mn^{2+}-holes}$.
Previous estimates of the valence band  quasiparticles  lifetimes  using  Fermi's
golden rule induced by this term
are of the order of 100-250 meV. \cite{Jungwirth:2002_c} We treat
$H_{\rm Mn^{2+}-holes}$ within the Born approximation in the Kubo
linear response formalism as in Ref. \onlinecite{Sinova:2002_,
Jungwirth:2002_c,Lopez-Sancho:2003_}. Hence, in what follows, we
drop $H_{\rm Mn^{2+}-holes}$ and introduce its effects through a
finite quasiparticle lifetime.

In order to obtain the temperature dependence of the electronic
and magnetic properties of the system, we minimize the free energy per unit volume,
\begin{equation} \label{Freeener}
\mathcal{F}= \mathcal{F}_{ions}+\mathcal{F}_{holes}\, \, \, \, \, ,
\end{equation}
where $\mathcal{F}_{ions}$ is the contribution of the ion spins to
the free energy and in the mean  field description has the form,
\begin{equation}
\mathcal{F}_{ions}= -T \, N_{Mn^{2+}} \,  \ln { \frac{ \sinh{(hS/T)}}
{(hS/2T)}} \, \, \, \, \label{freeiones}
\end{equation}
where $h=J p\xi/2$, $p$ and $N_{\rm Mn^{2+}}$ are the free carrier density
and Mn density respectively, and $\xi$ is the spin polarization of the carriers.
$\mathcal{F}_{holes}$ is the free energy of the holes, which is
obtained in the virtual crystal approximation (VCA) using a
Luttinger $\mathbf{k}\cdot \mathbf{p}$ model for describing the
carriers. In the VCA the average density of states for the real
system is replaced by that of the average Hamiltonian. This
approach implies a translational invariant system with an
effective exchange field acting on the carrier spins
$h_{ex}= m N_{Mn^{2+}} S  J$,
where $m$ is the polarization of the Mn spins. In  the
Luttinger $\mathbf{k}\cdot \mathbf{p}$ model the wave function of
the holes in the state $(n,\mathbf{k})$, where $n$ is the subband
index and $\mathbf{k}$ is the wave vector, is expressed as
\begin{equation}\label{wavefunction}
\psi_{n,\mathbf{k}} (\mathbf{r}) = e ^{i \mathbf{k} \cdot \mathbf{r
}} \sum_{J,m_j} \alpha_{n,\mathbf{k}} ^{J,m_J} |J,m_J \rangle \, \, \, \,
,
\end{equation}
where $|J,m_J\rangle$ are the six $\Gamma _{4v}$ valence band wave
functions. The coefficients $\alpha_{n,\mathbf{k}}^{J,m_J}$ and
the corresponding eigenvalues, $\varepsilon _ {n,\mathbf{k}}$,
depend on the spin polarization of the Mn, and are obtained, from
the Luttinger $\mathbf{k}\cdot \mathbf{p}$
Hamiltonian. \cite{Dietl:2001_b,Abolfath:2001_b}

By minimizing Eq.(\ref{Freeener}) we obtain the $T$-dependence of
the spin polarization of the carriers, $\xi (T)$ and  of the Mn,
$m (T)$. In the VCA the Curie temperature  has the expression $
T_c = N_{\rm Mn^{2+}} S^2J^{2} /3 \chi _p$ where $\chi _p$ is the
zero wavevector paramagnetic susceptibility of the hole gas.
Through the dependence of the Hamiltonian on the spin
polarization, we obtain the temperature dependence of the
eigenvectors, eigenvalues and chemical potential.  Throughout this
paper we consider always the typical optimal Mn concentration of
$x=0.05$ and an exchange coupling $J=60 \,{\rm
eVnm}^3$.\cite{Dietl:2000_,Okabayashi:1998_} Although the
properties of the system depend on the orientation of the
magnetization, this dependence is much smaller than the
T-dependence and since  we are interested in the variation of the
electronic properties with $T$ we fix the Mn spin polarization in
the $z$-direction. Experimentally this can be easily achieved by a
small coercive field of the order of 100 Oz.

From the eigenvalues and eigenvectors, we calculate the dielectric
function that in the RPA has the form,
\begin{equation}\label{epsilon}
    \frac{\epsilon(\mathbf{q},\omega)}{\epsilon_\infty} = 1 +
    \frac{4 \pi e ^2}{\epsilon_\infty q ^2 } \, \chi
    (\mathbf{q},\omega)  \, \, \, ,
\end{equation}
being $\chi (\mathbf{q},\omega)$ the susceptibility,
\begin{equation}\label{susceptibility}
\chi(\mathbf{q},\omega)  =   \sum _{i,j,\mathbf{k}}
\frac{n_F(\varepsilon _{i, \mathbf{k}+\mathbf{q}})
-n_F(\varepsilon _{j, \mathbf{k}}) }{\hbar \omega+\varepsilon _{j, \mathbf{k}}-
\varepsilon _{i,\mathbf{k}+\mathbf{q}}}
f_{i,j}(\mathbf{k},\mathbf{k}+\mathbf{q})  \, \, \, \, \, \ ,
\end{equation}
with
\begin{equation}\label{solape}
f_{i,j}(\mathbf{k},\mathbf{k}')  =  \left (  \sum _{J,m_J}
(\alpha_{i,\mathbf{k}}^{J,m_J})^* \,\alpha_{j,\mathbf{k}'
}^{J,m_J}  \right ) ^2 \, ,
\end{equation}
where $n_F $ denotes the Fermi-Dirac distribution and $\epsilon_\infty$
the dielectric function of the host semiconductor
($\epsilon_\infty$=10.90 for GaAs). In  the limit
$q\equiv|\mathbf{q}|\rightarrow 0$, the overlap function becomes
$f_{i,j}(\mathbf{k},\mathbf{k}+\mathbf{q})\rightarrow
\delta_{ij}+A_{i,j}(\mathbf{k})q^2$,  and Eq.~(\ref{epsilon}) can
be rewritten as:
\begin{equation}\label{epsilon2}
    \frac{\epsilon(\omega)}{\epsilon_\infty} = 1 -
\frac{\tilde{\omega}_{p}^2}{\omega^2}
    +\frac{4 \pi e ^2}{\epsilon_\infty} \, \chi_{inter}
    (\omega)  \, \, \, .
\end{equation}
The second term in the right hand side of Eq.~(\ref{epsilon2}) is the intra-band contribution and constitutes the
Drude weight present in dc measurements and it is closely related to the
plasmon frequency in the metallic regime,
\begin{equation}\label{omegaplasmon}
\tilde{\omega}_{p}^2\equiv\frac{4 \pi e
^2}{\epsilon_\infty}\sum_{i,\mathbf{k}} \delta (\varepsilon _{i,
\mathbf{k}}-\varepsilon _F) \frac{\partial \varepsilon _{i,
\mathbf{k}}}{\partial k}.
\end{equation}
The third term is the  inter-band contribution and has the form,
\begin{equation}\label{susceptibility2}
\chi_{inter}(\omega)  =   \sum _{i\neq j,\mathbf{k}}
\frac{n_F(\varepsilon _{i, \mathbf{k}}) -n_F(\varepsilon _{j,
\mathbf{k}}) }{\hbar \omega+\varepsilon _{j, \mathbf{k}}-
\varepsilon _{i,\mathbf{k}}} A_{i,j}(\mathbf{k})  \, \, \, \, \,.
\end{equation}
We also consider the coupling of the holes with longitudinal optical (LO) phonons.
Within the RPA approximation this can be done by including a phonon term in the
susceptibility such that the final dielectric function is:
\begin{equation}
\frac{\epsilon(\omega)}{\epsilon_\infty} = 1 -
\frac{\tilde{\omega}_{p}^2}{\omega^2} +\frac{4 \pi e
^2}{\epsilon_\infty} \, \chi_{inter} (\omega)
+\frac{\omega_{TO}^2-\omega_{LO}^2}
{\omega^2-\omega_{TO}^2+i\omega \gamma_{phonon}} \label{epsilon3}
\end{equation}
where $\hbar \omega_{TO}=33.25 {\rm
meV}$, $\hbar \omega_{LO}=36.23 {\rm meV}$. In the case of the
coupled plasmon-phonon systems, the damping of the optical phonons
is small compared to that of plasmons,\cite{Abstreiter}  and it is
estimated to be  $\gamma_{phonon}/\omega_{TO}\approx0.01$ for
GaAs. We account for disorder by including the lifetime broadening
of quasiparticle spectral functions in a phenomenological  way. We
use different lifetimes for the inter and intra-band contributions
to the dielectric function.
In general we expect  the intraband lifetime $1/\gamma _{ii}$ to
be much smaller than the interband lifetime $1/\gamma _{ij}$ as
the intraband scattering occurs at zero energy whereas the
interband scattering occurs at finite energy. This has been
demonstrated by finite size exact diagonalization studies where a
much larger intra-band disorder reduction was observed than the
interband contribution at finite frequencies.\cite{Yang:2003_} In
the previous formalism we have considered only the diagonal
component of the dielectric function parallel to the carrier spin
polarization.

\section{Results}
\label{results}
\subsection{Dielectric function}
\begin{figure}
\includegraphics[clip,width=8cm]{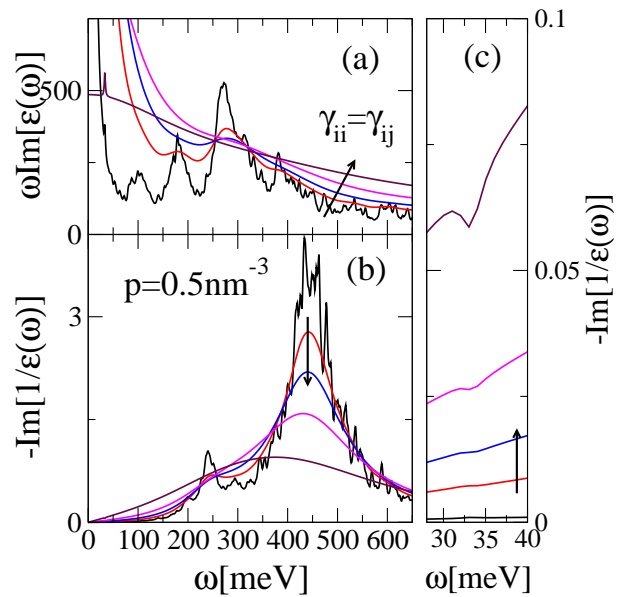}
\caption{(Color online.) Frequency dependence of the dielectric
function. a) $4\pi\sigma(\omega)=\omega {\rm
Im}[\epsilon(\omega)]$. b) $-{\rm
Im}[\frac{1}{\epsilon(\omega)}]$. c) Same as in b) showing a
blow-up of the coupled plasmon-phonon mode region. Parameters of
the calculation: $T=0 {\rm K}$, $p=0.5 {\rm nm}^{-3}$ and
different interband and intraband  scattering times:
$\gamma_{ii}=\gamma_{ij}$=2.5, 25, 50, 100 and 250meV.}
\label{fig1}
\end{figure}

\begin{figure}
\includegraphics[clip,width=8cm]{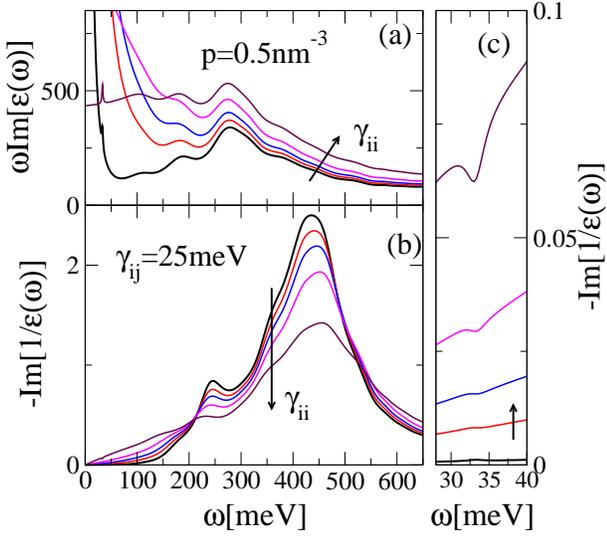}
\caption{(Color online.) Frequency dependence of the dielectric
function. a) $\omega {\rm
Im}[\epsilon(\omega)]=4\pi\sigma(\omega)$. b) $-{\rm
Im}[\frac{1}{\epsilon(\omega)}]$. c) Same as in b) showing a
blow-up of the coupled plasmon-phonon mode region. Parameters of
the calculation: $T=0 {\rm K}$, $p=0.5 {\rm nm}^{-3}$, fixed
$\gamma_{ij}$=25 meV and different $\gamma_{ii}$: 2.5, 25, 50, 100
and 250meV.}. \label{fig2}
\end{figure}

We have computed the dielectric function for different electron
densities, lifetimes and temperatures. We illustrate the
low-temperature behavior of the dielectric function and the ac
conductivity, $\sigma(\omega)$, in Fig.~\ref{fig1} where
$4\pi\sigma(\omega)=\omega {\rm Im}[\epsilon(\omega)]$ and $-{\rm
Im}[\frac{1}{\epsilon(\omega)}]$ are plotted for different degree
of disorder $\gamma=\gamma_{ii}=\gamma_{ij}$.
The conductivity, which is finite in the dc limit, has different
high-frequency contributions which can be explained in terms of
interband transitions: in the clean limit, at this carrier
density, there are four  strong resonances at $\omega\sim$ 100,
175, 275 and 390meV,
due to  transitions between the spin polarized six ${\bf k} \cdot
{\bf p}$ hole bands. Upon increasing $\gamma$, the Drude peak
contribution (intraband) broadens and overlaps with the
high-frequency interband contribution.  Our results agree with
those obtained previously by Sinova et al in
Ref.~\onlinecite{Sinova:2002_} and are consistent with the
experiments of Singley et al in Ref.~\onlinecite{Singley:2002_}
and ~\onlinecite{Singley:2003_}. Nonetheless, the dc conductivity
in the experiments is always much smaller than the ac conductivity
around $\omega\sim 2000cm^{-1}$. This is an indication that
intraband disorder is larger than interband disorder in the
experiments. This is also in agreement with finite size exact
diagonalization studies where a much larger intra-band spectral
weight reduction was observed due to disorder induced
localization.\cite{Yang:2003_} We illustrate this in
Fig.~\ref{fig2}a where we plot $4\pi\sigma(\omega)$ vs $\omega$
for fixed interband disorder $\gamma_{ij}$=25 meV and different
intraband disorder $\gamma_{ii}$. For $\gamma_{ii}>>\gamma_{ij}$
the dc limit is smaller than the ac conductivity for frequencies
up to $\omega\gtrsim 2000 \, {\rm cm}^{-1}$ (for $\gamma _{ii}$=250
meV) which corroborates our previous argument.

\begin{figure}
\includegraphics[clip,width=8cm]{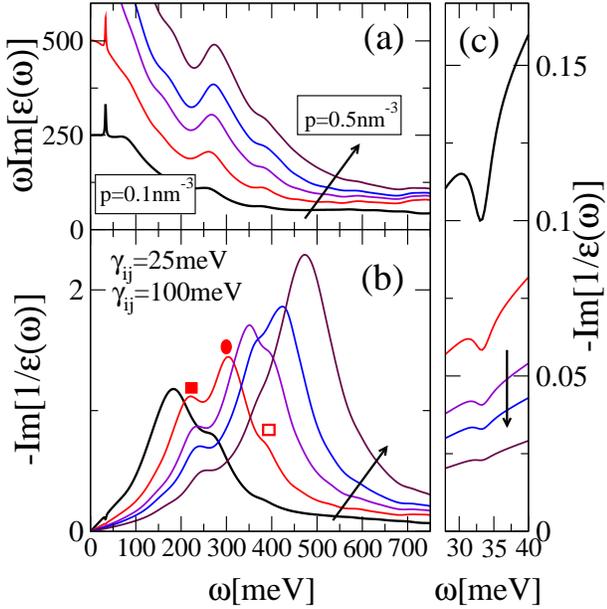}
\caption{(Color online) Frequency dependence of the dielectric
function. a) $\omega {\rm
Im}[\epsilon(\omega)]=4\pi\sigma(\omega)$. b) $-{\rm
Im}[\frac{1}{\epsilon(\omega)}]$. c) Same as in b) showing a
blow-up of the coupled plasmon-phonon mode region. Parameters of
the calculation: $T=0 {\rm K}$,  $\gamma_{ij}$=25 meV,
$\gamma_{ii}$=100meV, and hole densities varying from 0.1nm$^{-3}$
to  0.5nm$^{-3}$. The symbols mark the position of the
plasmon-like peak (circle) and of the peaks corresponding to
interband transitions (squares) for p=0.2nm$^{-3}$ (see text).}
\label{fig3}
\end{figure}

In Fig.~\ref{fig3} we plot $4 \pi {\rm Im}[\sigma(\omega)]$ and
$-{\rm Im} [1/\epsilon(\omega)]$ for different holes densities in
the case of zero temperature, $\gamma_{ii}$=100 meV, and
$\gamma_{ij}$=25 meV. As obtained by Sinova  {\it et al} in
Ref.~\onlinecite{Sinova:2002_}, we find that , as the hole density
increases a peak at energy near 280 meV emerges and the Drude peak
appears more clearly.

\begin{figure}
\includegraphics[clip,width=8cm]{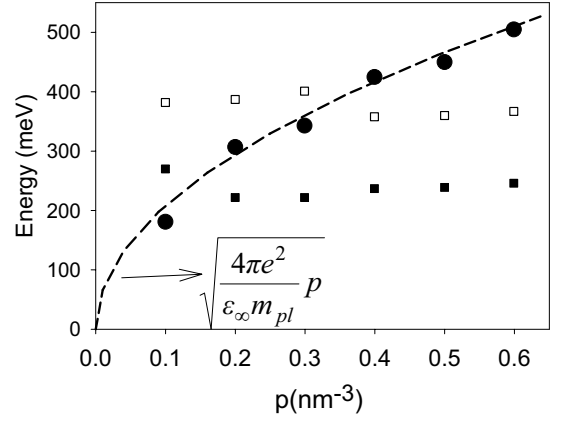}
\caption{Energy positions of the peaks appearing in $ {\rm
Im}[\frac{1}{\epsilon(\omega)}]$ as a function of the hole density
$p$. Circles corresponds to the stronger peak in the spectra
whereas square dot correspond to the weaker peaks. The dashed line
is the energy of a  plasmon  with an effective mass 0.25 times the
free electron mass.}  \label{fig4}
\end{figure}

The main features of $-{\rm
Im}[\frac{1}{\epsilon(\omega)}]$(Fig.~\ref{fig3}b) are a strong
single plasmon-like peak (marked in the figure with a circle for
p=0.2 nm$^{-3}$) with much weaker shoulders (marked with squares)
that correspond to interband transitions. The plasmon-like
character of the strong peak is evident when we plot the position
of the peaks in $-{\rm Im}[\frac{1}{\epsilon(\omega)}]$ as
function of the hole density (Fig.~\ref{fig4}). The energy of the
strong peaks (circles) can be fitted perfectly to the expression
$\sqrt{\frac{4 \pi e ^2 p}{\epsilon _{\infty}m_{pl}}}$, with
$m_{pl}$=0.25 $m_e$. The energies of the other peaks at
$\sim$230 meV and $\sim$375 meV (squares) depend very little on the
hole density, and we assign them to the light-hole band to
heavy-hole band  transition and to the light-hole band to the
split-off spin-orbit  band transition respectively. Of course
these transitions are coupled with the plasmon-like excitation.
Note that the peak positions are also rather independent of the
carrier spin polarization, as the concentration of Mn impurities
is fixed at x=5$\%$ for all the hole densities, and the carrier
polarization changes from 0.81 for $p$=0.1nm$^{-3}$ to 0.57 for
$p$=0.6nm$^{-3}$

\subsection{Sum rules}
\begin{figure}
\includegraphics[clip,width=8cm]{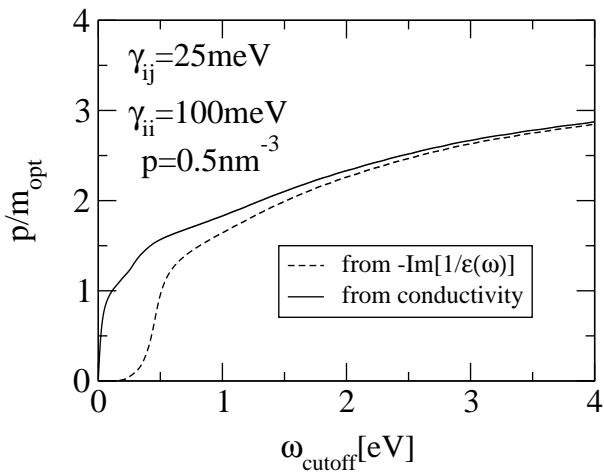}
\caption{Cut-off frequency dependence of the two sum rules. The
results correspond to $T=0 {\rm K}$, $p=0.5 {\rm nm}^{-3}$
,$\gamma_{ii}=100 {\rm meV}$ and $\gamma_{ij}=25 {\rm meV}$.}
\label{fig5}
\end{figure}

We can also define an optical  effective mass from the sum rules
of the model as
\begin{eqnarray}
\frac {p}{ m_{opt}} &\equiv& -\frac{2 \epsilon _{\infty}}{4 \pi ^2
e ^2}\int_0^{\infty} d\omega {\rm Im}[\frac{\epsilon
_{\infty}}{\epsilon(\omega)}]\nonumber \\ & =& \frac {2}{ \pi e
^2}\int_0^{\infty} d\omega {\rm Re}[\sigma(\omega)]
\label{sumrule} \end{eqnarray} Both sum rules, when integrated up
to infinity give the same optical mass. However, because the
calculation can only be done in a finite frequency range, in
practice the two sum rules  give different optical masses that
depend on the frequency cut-off in the integration, $\omega_{\rm cutoff}$.
In Fig.~\ref{fig5}, we plot the quantity $p/m_{opt}$ as
obtained from the two different sum rules as a function of
$\omega_{\rm cutoff}$ for $p$=0.5nm$^{-3}$. Note that it is
necessary to sum up to rather large values of the frequency in
order to get the same effective mass from the two different sum
rules. Also, we see that in the six-band ${\bf k} \cdot {\bf p}$
Hamiltonian it is necessary to integrate up to very high energies
for obtaining  an accurate value of $p/m_{\rm opt}$. That imposes a
serious limitation for  the experimental accurate estimation of this
quantity to a few percent but can still be used in combination
with other indirect transport measurements to establish consistency.

\begin{figure}
\includegraphics[clip,width=8cm]{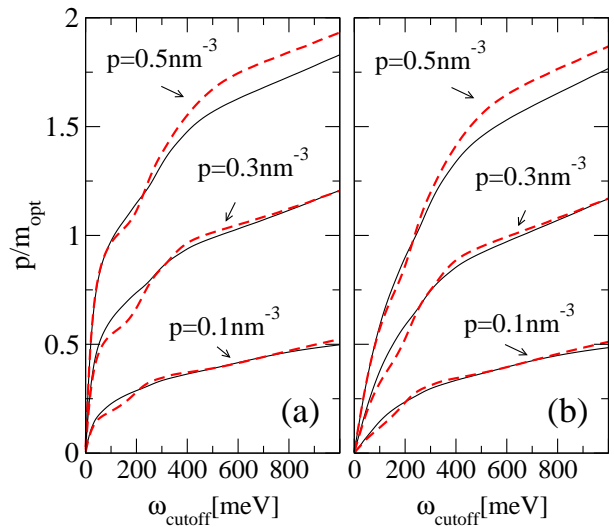}
\caption{(Color online) Cut-off frequency dependence of the sum
rule $\frac{p}{m_{opt}}\equiv\frac {2}{ \pi e
^2}\int_0^{\omega_{cutoff}} d\omega {\rm Re}[\sigma(\omega)]$. a)
$\gamma_{ii}=\gamma_{ij}=25 {\rm meV}$ for $T=0 {\rm K}$ (black
solid line) and $T=200 {\rm K}$ (red dashed line), for $p=0.1 {\rm
nm}^{-3}$ (bottom curves) and $p=0.5 {\rm nm}^{-3}$ (top curves)
b) Same as in a) but $\gamma_{ii}=250 {\rm meV}$ and
$\gamma_{ij}=25 {\rm meV}$.} \label{fig6}
\end{figure}

Recently, Singley {\it et al} \cite{Singley:2003_} have studied
experimentally the dependence of $p/m_{\rm opt}$ on the cutoff
frequency and temperature by integrating the measured infrared
conductivity up to 1.5 eV. In order to compare with these
experiments, we study $p/m_{\rm opt}$ as obtained from the
conductivity for different hole densities. The value of the
optical mass so obtained depends on the density of holes and, for
a finite frequency cutoff, on the lifetimes used in the
calculations. This is illustrated in Fig.\ref{fig6} where we plot
the conductivity sum rule as function of this frequency cut-off
for different temperatures, densities and quasiparticle lifetimes.
>From the figure we see that in the most metallic samples, lowest
$\gamma_{ii}$, the sum rule will be most accurate in determining
the carrier concentration. However, it is clear that for moderate
hole densities the value of $p/m_{opt}$ obtained from the sum rule
is far from the converged value. In any case the values of the
optical mass obtained from the sum rule are in the range
0.17-0.30$m_e$ depending on the density, frequency cutoff and
temperature. In Fig.\ref{fig6} we also compare the value of
$p/m_{opt}$ at zero temperature (ferromagnetic phase) with its
value at high temperature (paramagnetic phase). At low
temperatures there is an enhancement of the low energy spectral
weight. This can be attributed to low energy interband transitions
in the spin polarized system which originate from the spin-orbit
coupling. These transitions are absent in the paramagnetic phase
because there is not a Zeeman-like spin splitting of the bands. In
the absence of spin-orbit coupling, these excitations correspond
to transitions between bands with well defined spin antiparallel
polarization which, therefore, do not contribute to the
charge-charge response function, and are absent in the $ {\rm
Im}[\frac{1}{\epsilon(\omega)}]$ spectrum.

\subsection{Plasmon-phonon coupled modes}
For the range of carrier densities considered, a plasmon-phonon
coupled mode occurs at $\omega\approx \omega_{TO}$. This
plasmon-phonon coupled mode (Fig.~\ref{fig1}c) is very sensitive
to disorder: it changes from a peak-like shape to a Fano-like
shape upon increasing $\gamma_{ii}$. Interestingly, only intraband
disorder modifies significantly the coupled mode: $-{\rm
Im}[\frac{1}{\epsilon(\omega)}]$ around $\omega\approx
\omega_{TO}$ changes little when $\gamma_{ij}$ increases. This can
be seen by comparing the curves in Fig.~\ref{fig1}c (where
$\gamma_{ij}$ varies from 2.5meV to 500meV) with the corresponding
curves (i.e., with the same $\gamma_{ii}$) in Fig.~\ref{fig2}c
where $\gamma_{ij}$=25meV is fixed.  Due to the energy difference
between $\omega _{LO}$ and the interband transition energies, the
phonon modes almost do not couple with the interband transitions.
The plasmon-like part of the coupled mode is superimposed to the
plasmon peak for small $\gamma_{ii}$. On the other hand, for
$\gamma_{ii}\gtrsim 250$ meV the plasmon peak is strongly
overdamped (this corresponds to a phonon peak at
$\omega=\omega_{TO}$ superimposed to an almost featureless
$\sigma(\omega)$, see Fig.~\ref{fig1}a) and only the phonon-like
coupled mode remains. This is likely the situation corresponding
to the Raman experiments in Ref. \onlinecite{Seong:2002_} and
\onlinecite{Limmer:2002_}. Indeed, the authors of Ref.
\onlinecite{Limmer:2002_} model their data by using a simplified
Drude model where interband transitions are neglected, and obtain the value
of the carrier density indirectly by a line-shape
analysis of the coupled plasmon LO phonon in their Raman spectra.
By using an averaged hole mass, they conclude that the same
line-shape is obtained if one keeps the ratio
$\frac{p}{\gamma_{ii}}$ fixed, which allows them to estimate the
carrier density for different values of the plasmon damping.
Note, however, that in a simplified Drude model
$\varepsilon(\omega)\propto\frac{p}{m_{\rm opt}\gamma_{ii}}$ which,
together with our results, suggests that it should not be possible
to know $p$, $m_{\rm opt}$ and $\gamma_{ii}$ independently. Thus, we
believe that further experimental work is needed in order to
clarify this important point.

\section{Summary}
\label{summary} In closing, we have presented a theory of the
infrared dielectric function of (III,Mn)V ferromagnets based on
the kinetic exchange model for carrier induced (III,Mn)V
ferromagnetism. Our results demonstrate that the infrared
dielectric function in metallic samples can be used to obtain
information about the carrier concentration, to simulate the
plasmon-phonon overdamped coupled modes, and to explain naturally
the observed optical absorption measurements in the infrared
regime arising from inter-valence-band transitions. The predicted
$p/m_{\rm opt}$ ratios could be tested in the metallic regime
within the newly grown samples which exhibit large conductivities.
The utilization of the f-sum rule to estimate the carrier
concentration has been shown to be most accurate within the highly
metallic samples and should be tested experimentally by careful
comparison of these measurements with other means of measuring the
carrier concentration. We also have shown that the analysis of the
Raman plasmon-phonon coupled modes within a simple one band Drude
model as in Ref. \onlinecite{Limmer:2002_} are not very accurate
since estimates of  $m_{\rm opt}$ and $\gamma_{ii}$ cannot be done
independently.

\acknowledgements Financial support is acknowledged from Grants No
MAT2002-04429-C03-01, MAT2002-04095-C02-01 and MAT2002-02465
(MCyT, Spain) and Fundaci\'on Ram\'on Areces. Ram\'on Aguado also
acknowledges the support of the MCyT of Spain through the "Ram\'on
y Cajal" program. The authors acknowledge insightful conversations
with N. Samarth, D. N. Basov, and K. S. Burch.


\begin{thebibliography}{24}
\expandafter\ifx\csname
natexlab\endcsname\relax\def\natexlab#1{#1}\fi
\expandafter\ifx\csname bibnamefont\endcsname\relax
  \def\bibnamefont#1{#1}\fi
\expandafter\ifx\csname bibfnamefont\endcsname\relax
  \def\bibfnamefont#1{#1}\fi
\expandafter\ifx\csname citenamefont\endcsname\relax
  \def\citenamefont#1{#1}\fi
\expandafter\ifx\csname url\endcsname\relax
  \def\url#1{\texttt{#1}}\fi
\expandafter\ifx\csname
urlprefix\endcsname\relax\def\urlprefix{URL }\fi
\providecommand{\bibinfo}[2]{#2}
\providecommand{\eprint}[2][]{\url{#2}}

\bibitem[{\citenamefont{Ohno et~al.}(1996)\citenamefont{Ohno, Shen, Matsukura,
  Oiwa, Endo, Katsumoto, and Iye}}]{Ohno:1996_}
\bibinfo{author}{\bibfnamefont{H.}~\bibnamefont{Ohno}},
  \bibinfo{author}{\bibfnamefont{A.}~\bibnamefont{Shen}},
  \bibinfo{author}{\bibfnamefont{F.}~\bibnamefont{Matsukura}},
  \bibinfo{author}{\bibfnamefont{A.}~\bibnamefont{Oiwa}},
  \bibinfo{author}{\bibfnamefont{A.}~\bibnamefont{Endo}},
  \bibinfo{author}{\bibfnamefont{S.}~\bibnamefont{Katsumoto}},
  \bibnamefont{and} \bibinfo{author}{\bibfnamefont{Y.}~\bibnamefont{Iye}},
  \bibinfo{journal}{Appl. Phys. Lett.} \textbf{\bibinfo{volume}{69}},
  \bibinfo{pages}{363} (\bibinfo{year}{1996}).

\bibitem[{\citenamefont{Ohno}(1999)}]{Ohno:1999_a}
\bibinfo{author}{\bibfnamefont{H.}~\bibnamefont{Ohno}}, \bibinfo{journal}{J.
  Magn. Magn. Mater.} \textbf{\bibinfo{volume}{200}}, \bibinfo{pages}{110}
  (\bibinfo{year}{1999}).

\bibitem[{\citenamefont{Ku et~al.}(2003)\citenamefont{Ku, Potashnik, Wang,
  Seong, Johnston-Halperin, Meyers, Chun, Mascarenhas, Gossard, Awschalom
  et~al.}}]{Ku:2003_}
\bibinfo{author}{\bibfnamefont{K.~C.} \bibnamefont{Ku}},
  \bibinfo{author}{\bibfnamefont{S.~J.} \bibnamefont{Potashnik}},
  \bibinfo{author}{\bibfnamefont{R.~F.} \bibnamefont{Wang}},
  \bibinfo{author}{\bibfnamefont{M.~J.} \bibnamefont{Seong}},
  \bibinfo{author}{\bibfnamefont{E.}~\bibnamefont{Johnston-Halperin}},
  \bibinfo{author}{\bibfnamefont{R.~C.} \bibnamefont{Meyers}},
  \bibinfo{author}{\bibfnamefont{S.~H.} \bibnamefont{Chun}},
  \bibinfo{author}{\bibfnamefont{A.}~\bibnamefont{Mascarenhas}},
  \bibinfo{author}{\bibfnamefont{A.~C.} \bibnamefont{Gossard}},
  \bibinfo{author}{\bibfnamefont{D.~D.} \bibnamefont{Awschalom}},
  \bibnamefont{et~al.}, \bibinfo{journal}{Appl. Phys. Lett.}
  \textbf{\bibinfo{volume}{82}}, \bibinfo{pages}{2302} (\bibinfo{year}{2003}),
  \eprint{cond-mat/0210426}.

\bibitem[{\citenamefont{Yu et~al.}(2002)\citenamefont{Yu, Walukiewicz,
  Wojtowicz, Kuryliszyn, Liu, Sasaki, , and Furdyna}}]{Yu:2002_a}
\bibinfo{author}{\bibfnamefont{K.~M.} \bibnamefont{Yu}},
  \bibinfo{author}{\bibfnamefont{W.}~\bibnamefont{Walukiewicz}},
  \bibinfo{author}{\bibfnamefont{T.}~\bibnamefont{Wojtowicz}},
  \bibinfo{author}{\bibfnamefont{I.}~\bibnamefont{Kuryliszyn}},
  \bibinfo{author}{\bibfnamefont{X.}~\bibnamefont{Liu}},
  \bibinfo{author}{\bibfnamefont{Y.}~\bibnamefont{Sasaki}}, , \bibnamefont{and}
  \bibinfo{author}{\bibfnamefont{J.~K.} \bibnamefont{Furdyna}},
  \bibinfo{journal}{Phys. Rev. B} \textbf{\bibinfo{volume}{65}},
  \bibinfo{pages}{201303} (\bibinfo{year}{2002}).

\bibitem[{\citenamefont{Edmonds et~al.}(2004)\citenamefont{Edmonds,
  Boguslawski, Wang, Campion, Farley, Gallagher, Foxon, Sawicki, Dietl,
  Nardelli et~al.}}]{Edmonds:2004_}
\bibinfo{author}{\bibfnamefont{K.}~\bibnamefont{Edmonds}},
  \bibinfo{author}{\bibfnamefont{P.}~\bibnamefont{Boguslawski}},
  \bibinfo{author}{\bibfnamefont{K.}~\bibnamefont{Wang}},
  \bibinfo{author}{\bibfnamefont{R.}~\bibnamefont{Campion}},
  \bibinfo{author}{\bibfnamefont{N.}~\bibnamefont{Farley}},
  \bibinfo{author}{\bibfnamefont{B.}~\bibnamefont{Gallagher}},
  \bibinfo{author}{\bibfnamefont{C.}~\bibnamefont{Foxon}},
  \bibinfo{author}{\bibfnamefont{M.}~\bibnamefont{Sawicki}},
  \bibinfo{author}{\bibfnamefont{T.}~\bibnamefont{Dietl}},
  \bibinfo{author}{\bibfnamefont{M.}~\bibnamefont{Nardelli}},
  \bibnamefont{et~al.}, \bibinfo{journal}{Phys. Rev. Lett.}
  \textbf{\bibinfo{volume}{92}}, \bibinfo{pages}{037201}
  (\bibinfo{year}{2004}), \eprint{cond-mat/0307140}.

\bibitem[{\citenamefont{I.Kuryliszyn-Kudelska
  et~al.}(2003)\citenamefont{I.Kuryliszyn-Kudelska, T.Wojtowicz, X.Liu,
  J.K.Furdyna, W.Dobrowolski, J.Z.Domagala, E.Lusakowska, M.Goiran,
  E.Haanappel, and Portugall}}]{Kuryliszyn-Kudelska:2003_cond-mat/0304622}
\bibinfo{author}{\bibnamefont{I.Kuryliszyn-Kudelska}},
  \bibinfo{author}{\bibnamefont{T.Wojtowicz}},
  \bibinfo{author}{\bibnamefont{X.Liu}},
  \bibinfo{author}{\bibnamefont{J.K.Furdyna}},
  \bibinfo{author}{\bibnamefont{W.Dobrowolski}},
  \bibinfo{author}{\bibnamefont{J.Z.Domagala}},
  \bibinfo{author}{\bibnamefont{E.Lusakowska}},
  \bibinfo{author}{\bibnamefont{M.Goiran}},
  \bibinfo{author}{\bibnamefont{E.Haanappel}}, \bibnamefont{and}
  \bibinfo{author}{\bibfnamefont{O.}~\bibnamefont{Portugall}}
  (\bibinfo{year}{2003}), \eprint{cond-mat/0304622}.

\bibitem[{\citenamefont{Dietl et~al.}(2000)\citenamefont{Dietl, Ohno,
  Matsukura, Cibert, and Ferrand}}]{Dietl:2000_}
\bibinfo{author}{\bibfnamefont{T.}~\bibnamefont{Dietl}},
  \bibinfo{author}{\bibfnamefont{H.}~\bibnamefont{Ohno}},
  \bibinfo{author}{\bibfnamefont{F.}~\bibnamefont{Matsukura}},
  \bibinfo{author}{\bibfnamefont{J.}~\bibnamefont{Cibert}}, \bibnamefont{and}
  \bibinfo{author}{\bibfnamefont{D.}~\bibnamefont{Ferrand}},
  \bibinfo{journal}{Science} \textbf{\bibinfo{volume}{287}},
  \bibinfo{pages}{1019} (\bibinfo{year}{2000}).

\bibitem[{\citenamefont{Dietl et~al.}(2001)\citenamefont{Dietl, Ohno, and
  Matsukura}}]{Dietl:2001_b}
\bibinfo{author}{\bibfnamefont{T.}~\bibnamefont{Dietl}},
  \bibinfo{author}{\bibfnamefont{H.}~\bibnamefont{Ohno}}, \bibnamefont{and}
  \bibinfo{author}{\bibfnamefont{F.}~\bibnamefont{Matsukura}},
  \bibinfo{journal}{Phys. Rev. B} \textbf{\bibinfo{volume}{63}},
  \bibinfo{pages}{195205} (\bibinfo{year}{2001}).

\bibitem[{\citenamefont{Konig et~al.}(2003)\citenamefont{Konig, Schliemann,
  Jungwirth, and MacDonald}}]{Konig:2003_}
\bibinfo{author}{\bibfnamefont{J.}~\bibnamefont{Konig}},
  \bibinfo{author}{\bibfnamefont{J.}~\bibnamefont{Schliemann}},
  \bibinfo{author}{\bibfnamefont{T.}~\bibnamefont{Jungwirth}},
  \bibnamefont{and}
  \bibinfo{author}{\bibfnamefont{A.}~\bibnamefont{MacDonald}}, in
  \emph{\bibinfo{booktitle}{Electronic Structure and Magnetism of Complex
  Materials}}, edited by
  \bibinfo{editor}{\bibfnamefont{D.}~\bibnamefont{Singh}} \bibnamefont{and}
  \bibinfo{editor}{\bibfnamefont{D.}~\bibnamefont{Papaconstantopoulos}}
  (\bibinfo{publisher}{Springer Verlag Berlin}, \bibinfo{year}{2003}).

\bibitem[{\citenamefont{Linnarsson et~al.}(1997)\citenamefont{Linnarsson,
  Janzén, Monemar, Kleverman, and Thilderkvist}}]{Linnarsson:1997_}
\bibinfo{author}{\bibfnamefont{M.}~\bibnamefont{Linnarsson}},
  \bibinfo{author}{\bibfnamefont{E.}~\bibnamefont{Janzén}},
  \bibinfo{author}{\bibfnamefont{B.}~\bibnamefont{Monemar}},
  \bibinfo{author}{\bibfnamefont{M.}~\bibnamefont{Kleverman}},
  \bibnamefont{and}
  \bibinfo{author}{\bibfnamefont{A.}~\bibnamefont{Thilderkvist}},
  \bibinfo{journal}{Phys. Rev. B} \textbf{\bibinfo{volume}{55}},
  \bibinfo{pages}{6938} (\bibinfo{year}{1997}).

\bibitem[{\citenamefont{Yang et~al.}(2003)\citenamefont{Yang, Sinova,
  Jungwirth, Shim, and MacDonald}}]{Yang:2003_}
\bibinfo{author}{\bibfnamefont{S.-R.~E.} \bibnamefont{Yang}},
  \bibinfo{author}{\bibfnamefont{J.}~\bibnamefont{Sinova}},
  \bibinfo{author}{\bibfnamefont{T.}~\bibnamefont{Jungwirth}},
  \bibinfo{author}{\bibfnamefont{Y.}~\bibnamefont{Shim}}, \bibnamefont{and}
  \bibinfo{author}{\bibfnamefont{A.}~\bibnamefont{MacDonald}},
  \bibinfo{journal}{Phys. Rev. B} \textbf{\bibinfo{volume}{67}},
  \bibinfo{pages}{045205} (\bibinfo{year}{2003}), \eprint{cond-mat/0210149}.

\bibitem[{\citenamefont{Jungwirth et~al.}(2002)\citenamefont{Jungwirth,
  Abolfath, Sinova, Kucera, and MacDonald}}]{Jungwirth:2002_c}
\bibinfo{author}{\bibfnamefont{T.}~\bibnamefont{Jungwirth}},
  \bibinfo{author}{\bibfnamefont{M.}~\bibnamefont{Abolfath}},
  \bibinfo{author}{\bibfnamefont{J.}~\bibnamefont{Sinova}},
  \bibinfo{author}{\bibfnamefont{J.}~\bibnamefont{Kucera}}, \bibnamefont{and}
  \bibinfo{author}{\bibfnamefont{A.}~\bibnamefont{MacDonald}},
  \bibinfo{journal}{Appl. Phys. Lett.} \textbf{\bibinfo{volume}{81}},
  \bibinfo{pages}{4029} (\bibinfo{year}{2002}), \eprint{cond-mat/0206416}.

\bibitem[{\citenamefont{L\'opez-Sancho and Brey}(2003)}]{Lopez-Sancho:2003_}
\bibinfo{author}{\bibfnamefont{M.~P.} \bibnamefont{L\'opez-Sancho}}
  \bibnamefont{and} \bibinfo{author}{\bibfnamefont{L.}~\bibnamefont{Brey}},
  \bibinfo{journal}{Phys. Rev. B} \textbf{\bibinfo{volume}{68}},
  \bibinfo{pages}{113201} (\bibinfo{year}{2003}), \eprint{cond-mat/0302237}.

\bibitem[{not()}]{note_dassarma}
\bibinfo{note}{The study of optical conductivity in ferromagnetic
  semiconductors in the impurity band regime, has been addresed by Hwang {\it
  et al.} in reference \cite{Hwang:2002_}.}

\bibitem[{\citenamefont{Blinowski and Kacman}(2003)}]{Blinowski:2003_}
\bibinfo{author}{\bibfnamefont{J.}~\bibnamefont{Blinowski}} \bibnamefont{and}
  \bibinfo{author}{\bibfnamefont{P.}~\bibnamefont{Kacman}},
  \bibinfo{journal}{Phys. Rev. B} \textbf{\bibinfo{volume}{67}},
  \bibinfo{pages}{121204} (\bibinfo{year}{2003}), \eprint{cond-mat/0212093}.

\bibitem[{\citenamefont{Sinova et~al.}(2002)\citenamefont{Sinova, Jungwirth,
  Yang, Kucera, and MacDonald}}]{Sinova:2002_}
\bibinfo{author}{\bibfnamefont{J.}~\bibnamefont{Sinova}},
  \bibinfo{author}{\bibfnamefont{T.}~\bibnamefont{Jungwirth}},
  \bibinfo{author}{\bibfnamefont{S.-R.~E.} \bibnamefont{Yang}},
  \bibinfo{author}{\bibfnamefont{J.}~\bibnamefont{Kucera}}, \bibnamefont{and}
  \bibinfo{author}{\bibfnamefont{A.}~\bibnamefont{MacDonald}},
  \bibinfo{journal}{Phys. Rev. B} \textbf{\bibinfo{volume}{66}},
  \bibinfo{pages}{041202} (\bibinfo{year}{2002}), \eprint{cond-mat/0204209}.

\bibitem[{\citenamefont{Abolfath et~al.}(2001)\citenamefont{Abolfath,
  Jungwirth, Brum, and MacDonald}}]{Abolfath:2001_b}
\bibinfo{author}{\bibfnamefont{M.}~\bibnamefont{Abolfath}},
  \bibinfo{author}{\bibfnamefont{T.}~\bibnamefont{Jungwirth}},
  \bibinfo{author}{\bibfnamefont{J.}~\bibnamefont{Brum}}, \bibnamefont{and}
  \bibinfo{author}{\bibfnamefont{A.}~\bibnamefont{MacDonald}},
  \bibinfo{journal}{Phys. Rev. B} \textbf{\bibinfo{volume}{63}},
  \bibinfo{pages}{054418} (\bibinfo{year}{2001}).

\bibitem[{\citenamefont{Okabayashi et~al.}(1998)\citenamefont{Okabayashi,
  Kimura, Rader, Mizokawa, Fujimori, Hayashi, and Tanaka}}]{Okabayashi:1998_}
\bibinfo{author}{\bibfnamefont{J.}~\bibnamefont{Okabayashi}},
  \bibinfo{author}{\bibfnamefont{A.}~\bibnamefont{Kimura}},
  \bibinfo{author}{\bibfnamefont{O.}~\bibnamefont{Rader}},
  \bibinfo{author}{\bibfnamefont{T.}~\bibnamefont{Mizokawa}},
  \bibinfo{author}{\bibfnamefont{A.}~\bibnamefont{Fujimori}},
  \bibinfo{author}{\bibfnamefont{T.}~\bibnamefont{Hayashi}}, \bibnamefont{and}
  \bibinfo{author}{\bibfnamefont{M.}~\bibnamefont{Tanaka}},
  \bibinfo{journal}{Phys. Rev. B} \textbf{\bibinfo{volume}{58}},
  \bibinfo{pages}{R4211} (\bibinfo{year}{1998}).

\bibitem[{\citenamefont{G.Abstreiter et~al.}(1984)\citenamefont{G.Abstreiter,
  M.Cardona, and A.Pinczuk}}]{Abstreiter}
\bibinfo{author}{\bibnamefont{G.Abstreiter}},
  \bibinfo{author}{\bibnamefont{M.Cardona}}, \bibnamefont{and}
  \bibinfo{author}{\bibnamefont{A.Pinczuk}}, in \emph{\bibinfo{booktitle}{Light
  Scattering in Solids IV}}, edited by
  \bibinfo{editor}{\bibnamefont{M.Cardona}} \bibnamefont{and}
  \bibinfo{editor}{\bibnamefont{G.Guntherodt}}
  (\bibinfo{publisher}{Springer-Verlag, Berlin}, \bibinfo{year}{1984}).

\bibitem[{\citenamefont{Singley et~al.}(2002)\citenamefont{Singley, Kawakami,
  Awschalom, and Basov}}]{Singley:2002_}
\bibinfo{author}{\bibfnamefont{E.~J.} \bibnamefont{Singley}},
  \bibinfo{author}{\bibfnamefont{R.}~\bibnamefont{Kawakami}},
  \bibinfo{author}{\bibfnamefont{D.~D.} \bibnamefont{Awschalom}},
  \bibnamefont{and} \bibinfo{author}{\bibfnamefont{D.~N.} \bibnamefont{Basov}},
  \bibinfo{journal}{Phys. Rev. Lett.} \textbf{\bibinfo{volume}{89}},
  \bibinfo{pages}{097203} (\bibinfo{year}{2002}).

\bibitem[{\citenamefont{Singley et~al.}(2003)\citenamefont{Singley, Burch,
  Kawakami, Stephens, Awschalom, and Basov}}]{Singley:2003_}
\bibinfo{author}{\bibfnamefont{E.~J.} \bibnamefont{Singley}},
  \bibinfo{author}{\bibfnamefont{K.~S.} \bibnamefont{Burch}},
  \bibinfo{author}{\bibfnamefont{R.}~\bibnamefont{Kawakami}},
  \bibinfo{author}{\bibfnamefont{J.}~\bibnamefont{Stephens}},
  \bibinfo{author}{\bibfnamefont{D.~D.} \bibnamefont{Awschalom}},
  \bibnamefont{and} \bibinfo{author}{\bibfnamefont{D.~N.} \bibnamefont{Basov}},
  \bibinfo{journal}{Phys. Rev. B} \textbf{\bibinfo{volume}{68}},
  \bibinfo{pages}{165204} (\bibinfo{year}{2003}).

\bibitem[{\citenamefont{Seong et~al.}(2002)\citenamefont{Seong, Chun, Cheong,
  Samarth, and Mascarenhas}}]{Seong:2002_}
\bibinfo{author}{\bibfnamefont{M.~J.} \bibnamefont{Seong}},
  \bibinfo{author}{\bibfnamefont{S.~H.} \bibnamefont{Chun}},
  \bibinfo{author}{\bibfnamefont{H.~M.} \bibnamefont{Cheong}},
  \bibinfo{author}{\bibfnamefont{N.}~\bibnamefont{Samarth}}, \bibnamefont{and}
  \bibinfo{author}{\bibfnamefont{A.}~\bibnamefont{Mascarenhas}},
  \bibinfo{journal}{Phys. Rev. B} \textbf{\bibinfo{volume}{66}},
  \bibinfo{pages}{033202} (\bibinfo{year}{2002}).

\bibitem[{\citenamefont{Limmer et~al.}(2002)\citenamefont{Limmer, Glunk,
  Mascheck, Koeder, Scoch, Thonke, Sauer, and Waag}}]{Limmer:2002_}
\bibinfo{author}{\bibfnamefont{W.}~\bibnamefont{Limmer}},
  \bibinfo{author}{\bibfnamefont{M.}~\bibnamefont{Glunk}},
  \bibinfo{author}{\bibfnamefont{S.}~\bibnamefont{Mascheck}},
  \bibinfo{author}{\bibfnamefont{A.}~\bibnamefont{Koeder}},
  \bibinfo{author}{\bibfnamefont{W.}~\bibnamefont{Scoch}},
  \bibinfo{author}{\bibfnamefont{K.}~\bibnamefont{Thonke}},
  \bibinfo{author}{\bibfnamefont{R.}~\bibnamefont{Sauer}}, \bibnamefont{and}
  \bibinfo{author}{\bibfnamefont{A.}~\bibnamefont{Waag}},
  \bibinfo{journal}{Phys. Rev. B} \textbf{\bibinfo{volume}{66}},
  \bibinfo{pages}{205209} (\bibinfo{year}{2002}).

\bibitem[{\citenamefont{Hwang et~al.}(2002)\citenamefont{Hwang, Millis, and
  Sarma}}]{Hwang:2002_}
\bibinfo{author}{\bibfnamefont{E.~H.} \bibnamefont{Hwang}},
  \bibinfo{author}{\bibfnamefont{A.~J.} \bibnamefont{Millis}},
  \bibnamefont{and} \bibinfo{author}{\bibfnamefont{S.~D.} \bibnamefont{Sarma}},
  \bibinfo{journal}{Phys. Rev. B} \textbf{\bibinfo{volume}{65}},
  \bibinfo{pages}{233206} (\bibinfo{year}{2002}).

\end{thebibliography}

\end{document}